# Comments on "A model-based design methodology for the development of mechatronic systems"

**Kleanthis Thramboulidis**
Electrical and Computer Engineering, University of Patras, Greece.

**A B S T R A C T**
In the paper by G. *Barbieri et al.* (*Mechatronics (2014), http://dx.doi.org/10.1016/j.mechatronics. 2013.12.004*), a design methodology, based on the W life cycle process model, is presented and SysML is proposed as a tool to support the whole development process. In this letter, we discuss the presented approach, we point out technical errors and raise additional issues that might help in making the proposed approach applicable.

## 1. Introduction

The main contribution of [1], as claimed by the authors, is the introduction of a design methodology based on the W model and the identification of SysML as the tool to support the whole process. To this direction, in Section 3 they describe the "different phases" of their "integrated design methodology", which is based on the W model that captures the five of these. Then, in Section 4, they present the process of testing the presented methodology in an industrial case study, i.e., the SpA filling machine.

In this letter, we discuss the presented methodology and its application to the case study; we point out unjustifiable claims made by the authors of [1], inconsistencies in the models and technical errors that make the presented methodology not easy to follow by the developer. To begin with, while the authors claim to be presenting a Model-based methodology, they do not address the lack of tool interoperability, which is considered "a significant inhibitor to widespread deployment of MBSE" [1, Ref. 7]. Also, their initial claim of using SysML to construct the system model is not followed when in [1, Sec. 4.5] the control system is designed in Simulink and in [1, Sec. 4.3] the AMESim model is developed during the virtual system integration phase. Moreover, the authors just use hyperlinks for connecting the SysML models with the domain specific models of the various disciplines [1, Sec. 2.3.3]. This is in contrast with their own claim that "the information exchange among the SysML model and DSMs" is one of the fundamental aspects to realize the goal of integrated design [1, Sec. 3.3]. Authors propose to decompose the system into modules, and the modules into components [1, Sec. 3], but they not argue on this three level composition hierarchy and they do not provide the definition for the module and component as well as their difference. Finally, it is not clear if these artifacts, i.e., module and component, are mono-discipline or multi-discipline ones.

This letter is organized as follows: In Section 2 we discuss issues related to the system life cycle process model and we argue that the critique expressed in [1] on the V model is unjustifiable. In Section 3, the contribution of [1] related to SysML is discussed. Finally, in Section 4 we comment on the industrial case study being used in [1] to test the presented methodology. Our intention is to clarify several open issues and give the authors the ability to argue on the presented approach.

## 2. The System Life Cycle Process Model

The term process or software life-cycle model was used to refer to models of software evolution, which appear with the first large projects developing software systems. These models provide an abstract scheme accounting for the "natural" or engineered development of software systems [2]. The term was later used in system engineering (SE) and the International Council on Systems Engineering's (INCOSE) identifies it as one of SE's key concepts. INCOSE defines the life cycle for a system as consisting of "a series of stages regulated by a set of management decisions which confirm that the system is mature enough to leave one stage and enter another" [3].

Authors in [1] claim that the V model imposes the design of mechatronic systems to "be separated into the development of single components, which should be designed parallel in the single disciplines and then be integrated to the overall system." They also claim that according to the V model "the modules would be designed parallel until all the components were completely developed. Then, the control system would be introduced in order to cover the gaps of the mechanics." Moreover, authors claim that in order to overcome the problems of late integration, which is imposed by the V model, they have adopted the W model that provides a virtual system integration. However, these claims are completely arbitrary.

1. INCOSE characterizes the V Model (Vee Model) as the primary example of pre-specified and sequential processes but it admits that a) some of its variations are considered as primarily evolutionary and concurrent processes, and b) both the initial V model and its variations address the basic set of SE activities. It should be noted that the V model enables, according to INCOSE, the developers to perform concurrent opportunity and risk analyses, as well as continuous in-process validation.
2. The VDI guideline 2206, entitled "Design methodology for mechatronic systems" [5], which has been developed by VDI, the German engineers association, defines the domain-specific design, i.e., mechanical, electrical and information engineering as a concurrent process at the bottom of the V-model, where it also assigns the model analysis activities.
3. The V model has been used in a synergistic integration process, at the component level, of the three discipline constituent parts of the Mechatronic system, i.e., mechanics, electronics and software [4]. Based on this, the system is considered as a composition of mechatronic components, which are specified at the left side of the V model and are developed at the bottom of the V model following a concurrent synergistic engineering process for the three discipline constituent parts of the mechatronic component. It should be noted that the system level model, which is expressed in SysML, is analyzed through simulation and since there is no SysML engine at the time to execute the model, existing model analysis engines from the three disciplines are used for this purpose.
4. The V model and modified versions of its "have been applied extensively in the areas of systems engineering and systems development" [10].

*2.1. System integration*

The author's claim that the "late integration would likely determine consistency problems among the different modules that were designed independently and the generation of a non-optimal solution" [1, Sec. 3.3] is misleading. It is obvious that they refer to the system integration process that is captured on the right side of the V model and represents the integration process of the implemented artifacts. However, it should be noted that the integration of the system components has already been considered at the model level during the system modeling process (system architecture definition) and the specifications of the constituent components have been defined in such a way as to effectively address the integration problem. The independent development of components, which is carried out after their specification at the system level, concerns only their implementation which should be fully consistent with their specifications defined at the system architecture model [4]. This consistency is verified during the unit test process, while the consistency of the integration is verified during the system integration phase. Verification is a continuous activity during the system modeling process.

In the virtual system integration phase, authors claim that they built the AMESim model [1, Sec. 4.3] without giving any information on how is this related to the already constructed SysML system model. Moreover, they note, completely arbitrary, that this phase "allows the parallel development of the mechanics and the control devices of the system" and for this reason it "constitutes a fundamental step for an integrated design." [1, Sec. 3.3].

Authors give their own definition for the system integration phase without arguing on this. System integration is a well known and widely accepted and used term in software and system engineering. INCOSE in its guide to the System Engineering Body of Knowledge defines this phase as "taking delivery of the implemented system elements which compose the system-of-interest (SoI), assembling these implemented elements together, and performing the verification and validation actions (V&V actions) in the course of the assembly. The ultimate goal of system integration is to ensure that the individual system elements function properly as a whole and satisfy the design properties or characteristics of the system." However, authors claim that they start the system integration with the identification of the system behavior and that at the phase of system integration they had just defined which operations must be performed inside the states, but they had "not specified the method yet" [1, Sec. 4.5]. Moreover, they define the system interfaces after the definition of the system behavior [1, Sec. 3.5].

**3. The System Modeling Language (SysML)**

Regarding SysML, authors claim that: a) "Up to date, SysML has been mostly used as tool for reverse engineering processes," and b) one of the main contributions of their paper is "the identification of SysML as the tool to support the whole (development) process", and "making SysML the main tool for the complete design of mechatronic systems through an integrated approach."

However, SysML has been defined as a language for system engineering and it is widely used for this purpose for several years now, e.g., [4][11][12]. SysML is the result of the INCOSE strategic decision to customize the Unified Modeling Language (UML) for systems engineering applications [6]. The intention was to unify the diverse modeling languages currently used by systems engineers. Moreover, as claimed in [10], "several commercially-offered model-based systems and software engineering methodologies, including most of the MBSE methodologies surveyed by this study, incorporate the UML and/or SysML into specific methods and artifacts produced as part of the methodology."

Authors also claim that SysML is object-oriented. However, as it is emphasized in [6]: a) SysML is "intended to support multiple processes and methods such as structured, object-oriented, and others," and b) is the methodology that imposes additional constraints on how a construct or diagram kind may be used. The above are considered as advantages of the language compared to UML, which is a language for object-oriented development. UML was widely used in SE before the definition of SysML. A survey of applications of UML in mechatronic systems is given in [9].

What authors call "external system diagram" is well known in software and system engineering as context diagram and can be represented by using use case diagram or internal block diagram [6]. It should be noted that flow ports used by authors in the "external system diagram" are deprecated [6].

## 4. The Case Study

Authors use as industrial case study the filling system of a Tetra Pak Packaging Solution SpA filling machine, which is shown in [1, Fig. 6]. However, the legend of this figure informs the reader that "The filling system is not represented." Our first comment in this sub section is that the main artifact of a Model-based approach is a coherent model of the system being developed [13]. However, the presented in [1] model of the case study is far from been coherent. We argue on this in the following sub-sections.

### 4.1. Context Diagram

From the context diagram (external system diagram for the authors) given in Fig. 1 ([1, Fig. A.8]), which represents in green the SoI, i.e., the filling system, it is obvious that authors consider the PLC, and subsequently the software that is executed on it as an external entity that interact with the filling system with a connector through which information is exchanged. This means that the SoI, shown as part *:Filling system*, represents just the mechanical unit filling system, of the Tetra Pak filling machine.

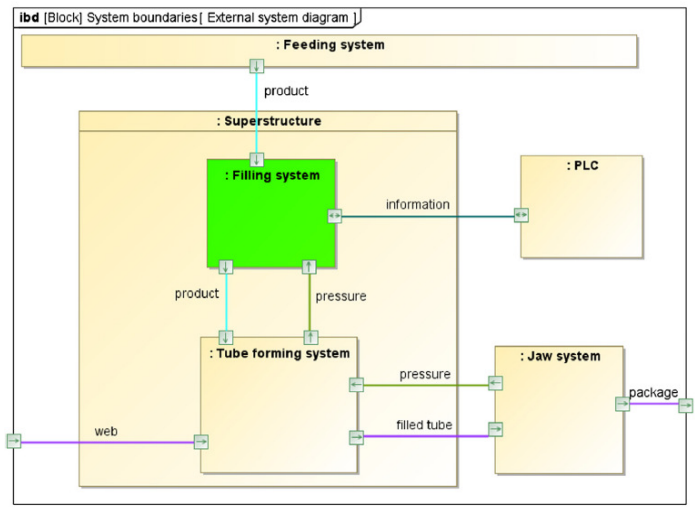

Fig. 1. The context diagram of the SoI [1].

Fig. 2a ([1, Fig. 6]) represents, taking into account also Fig. 2b ([14, Fig. 2) and Fig. 2c ([14, Fig. 3), the mechanical part of the Tetra Pak filling machine and in this sense the PLC and the controlling software that is executed on it are considered as external entities as captured in the context diagram of Fig. 1 ([1, Fig. A.8]). Based on Fig. 2, the mechanical part of the Tetra Pak filling machine is composed of a part that is called superstructure, which interacts with the other part that is called Jaw system. The superstructure is composed of the Tube forming section and another part that is not named. It is a question why the filling system is not shown in this figure.

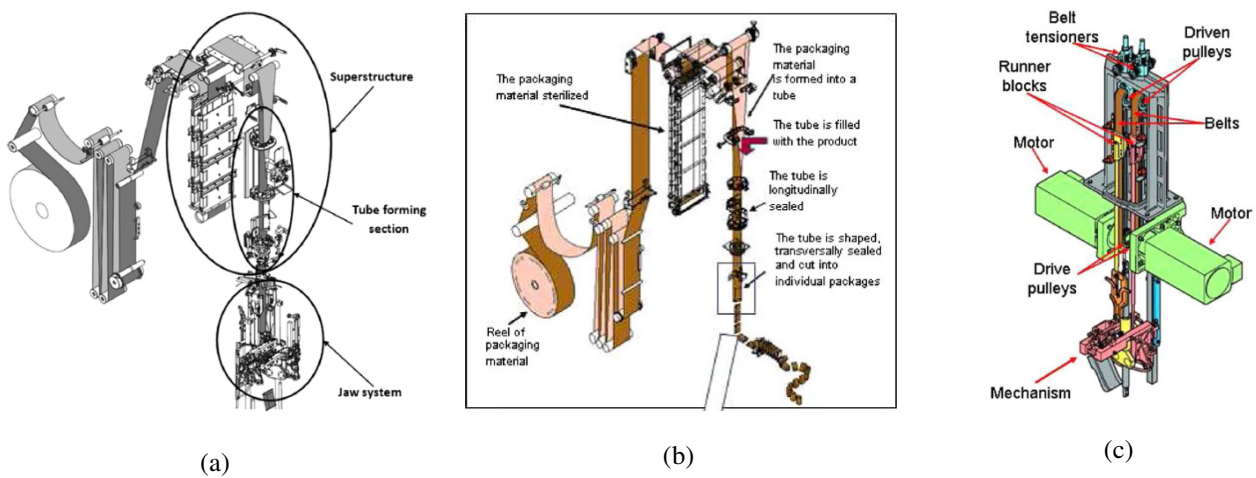

Fig. 2. The Tetra Pak filling machine part of which is the SoI [1][14].

According to the context diagram of the SoI (Fig. 1), the superstructure is composed of the *:Filling system* and the *:Tube forming system,* which is not consistent with Fig. 2a. The SoI interacts with the part *:Tube forming system* through the *product* and *pressure* connectors. However, in the sequence diagram of Fig. 3 ([1, Fig. A.9]) the *:Filling system* block accepts *Liquid level feedback* from the *:Tube forming system* which is not consistent with Fig. 1. It should be noted that the PLC, that is assumed to execute the controlling software for the whole machine, does not interact with any other part of the filling machine, which raises many questions on the correctness of the model.

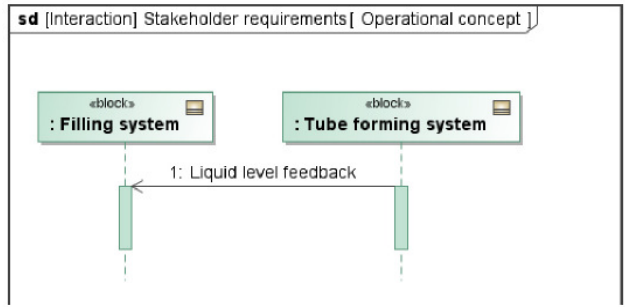

Fig. 3. Sequence diagram [1].

Authors use the term information to show that information flows through the connector connecting the SoI and the *:PLC*. It should be noted that information may be processes, transferred and stored, and this can be performed using either software, electronics or mechanics. Information may be transferred either by data exchange, by signal, or even by pressure or force. For example the connector between the *:Tube forming system* and the *:Jaw system* that is named pressure may represent just information transfer between the interacting systems. Based on this reasoning the naming but mainly the convention of the authors to use colors to increase the comprehension of the model is not only effectless but may lead to errors in modeling. It is also questionable why authors claim that electrical (interactions) are not specified in the context diagram but are defined after the detailed system design [1, Sec. 3.1.2].

Regarding the context diagram we consider the following two alternatives:
1. The SoI is the mechanical part of the filling System. In this case, the controller of the filling system, who's functionality will be implemented by software executed on a PLC, would be represented as an external entity in the context diagram. In this case, it is obvious that the SoI is not mechatronic and there is no need for a mechatronic design methodology.
2. The SoI is the whole filling system. In this case, the controller is part of the SoI and is erroneous to represent it as an external entity. The other captured external entities should represent the corresponding mechatronic systems [8] that interact with the SoI.

*4.2. Requirements handling through the development process*

Authors identify only one use case, i.e., the ''form packages with the correct volume of liquid'' but they do not refer which actor is initiating this use case. It is also not clear how this use case is realized through the sequence diagram of fig. A.9, as it is claimed by authors in [1, Sec. 4.1]. It should be noted that the package, i.e., the result of the use case, is produced by the *:Jaw system* as shown in the context diagram.

Authors next claim that they have defined the other stakeholder requirements and they broke them down for the generation of the system requirements. This raises the question of which requirements are captured using use cases and which in a different way. Since it seems that the four functions have resulted from the one use case, the question is what is the effect of the other requirements to the definition of the system functional architecture. Furthermore, the described process raises the question of what paradigm are the authors applying for the definition of the system architecture. The procedural one, according to which the system is considered as a composition of interacting processes, which implement the system functions, or the object-oriented one, according to which the system is considered as a composition of collaborating objects, i.e., modules? From the description it seems that they first apply the procedural paradigm to get the data flow diagram (DFD), which they represent by using the UML activity diagram notation, and then they transform this to an object-oriented architecture. The construction of both models, i.e., the procedural and the object-oriented, complicates the development process and authors have to argue on the benefits of such an approach.

Through the above process authors claim that they have identified four functions that have to be performed by the SoI, and assign these to an equal number of modules shown in Fig. 4 ([1, Fig. A.10]). It should be noted that the ibd of this figure is not consistent with the context diagram since the Filling system appears to have as input *liquid level feedback,* which is not shown in Fig. 1 ([1, Fig. A.8]). Moreover, *pressure* and *information flow,* shown in the context diagram, do not appear in the Filling system ibd.

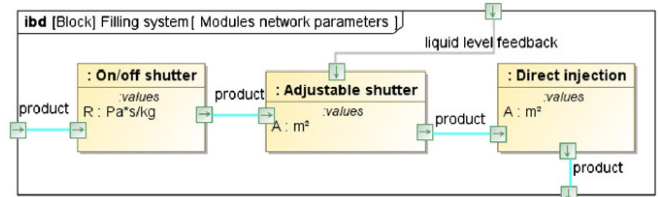

Fig. 4. Internal block diagram of the Filling system [1].

*4.3. Functional architecture*

Authors next claim that they apply the W model to the identified modules and they present in Fig. 5 ([1, Fig. A.11]) the functional architecture of the adjustable shutter module. Among the several questions which are raised on this process, we discriminate the following:

1. Is the *receive liquid level feedback* a function that has to be defined during the architecture definition process? If yes, then what is the difference between the architecture definition and the detailed design, two well known activities in software and system development?
2. What is the benefit of applying the W model for such low level modules? Is it effective to apply the W model to get the functional architecture of the module adjustable shutter?
3. Do we have to apply the W model during the development process in order to define activity diagrams that model behavior of modules?

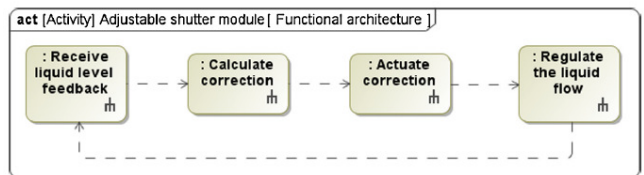

Fig. 5. Activity diagram of the Adjustable shutter module [1].

The following definition of the term methodology given by Booch et al. in [7] may help in answering the above questions. The methodology guides the engineer in deciding what artifacts to produce, what activities and what workers to use to create and manage them, and how to use those artifacts to measure and control the project as a whole. These guidelines are not defined by the process model which is an abstract representation of the basic phases. These phases do not define details on what, when, and how the various artifacts should be produced.

*4.4. System structure and behavior*

According to the context diagram given in Fig. 1, the *:Filling system* represents only the mechanical part of the Filling system. This means that the parts of the ibd of the Filling system shown in Fig. 4 represent also mechanical parts but this is not consistent with the functional architecture of the adjustable shutter presented in Fig. 5. A mechanical part, i.e., the *:Adjustable shutter*, appears to execute computational activities, such as *:Calculate correction*. Furthermore, it seems that the *:Receive liquid level feedback* accepts as input the *liquid level feedback* shown in the ibd of Fig. 4 but it is a question of how the output of the *:Calculate correction* activity affects the system since there is no related output at the corresponding ibd.

Authors after defining the functional architecture of the *adjustable shutter* they identify a servomotor as a component for actuating the correction and describe the behavior of this component by using the state machine shown in Fig. 6 ([1, Fig. A.12]). Based on this state chart, the servomotor performs just one operation and has only one state, the one that represents the rotation of the cam. It is not clear what exactly this state machine represents and which are the conditions that fire the transitions. Does it represent the states of the mechanical servomotor, the states of its controlling software entity or something else?

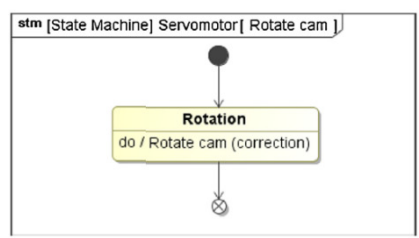

Fig. 6. Servomotor state machine [1].

Authors present in Fig. 7 ([1, Fig. A.13]) the adjustable shutter state machine. This state machine is incomplete, not consistent with the activity diagram of adjustable shutter given in Fig. 5 and is not correctly synchronized with the state machine of the Servomotor given in Fig. 6.

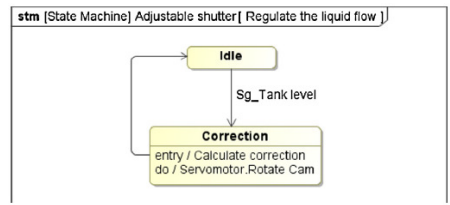

Fig. 7. The Adjustable shutter state machine [1]

**Acknowledgments**
This work was completed while the author was a visiting professor at TUM, partially funded by Bayerische Forschungsstiftung.

**References**
[1] Barbieri G. et al. A model-based design methodology for the development of mechatronic systems. Mechatronics (2014), http://dx.doi.org/10.1016/j.mechatronics.2013.12.004
[2] Scacchi. Walter, "Models of Software Evolution: Life Cycle and Process," Software Engineering Institute, Carnegie Mellon University, Pittsburgh, Pennsylvania, Curriculum Module CMU/SEI-87-CM-010, 1987. http://resources.sei.cmu.edu/library/asset-view.cfm?AssetID=10191
[3] INCOSE. 2012. *Systems Engineering Handbook: A Guide for System Life Cycle Processes and Activities*, version 3.2.2. San Diego, CA, USA: International Council on Systems Engineering (INCOSE), INCOSE-TP-2003-002-03.2.2.
[4] K. Thramboulidis, "The 3+1 SysML View-Model in Model Integrated Mechatronics", Journal of Software Engineering and Applications (JSEA), vol.3, no.2, 2010, pp.109-118
[5] http://www.vdi.de/uploads/tx_vdirili/pdf/9567281.pdf
[6] OMG, "OMG Systems Modeling Language (OMG SysML™)," Version 1.3, June 2012.
[7] Grady Booch, James Rumbaugh, and Ivar Jacobson, The unified modeling language user guide, Addison-Wesley, 1999.
[8] K. Thramboulidis, "A Framework for the Implementation of Industrial Automation Systems Based on PLCs", (submitted)
[9] Fernando Valles-Barajas, "A survey of UML applications in mechatronic systems", Innovations System Software Engineering (2011) 7:43–51 DOI 10.1007/s11334-011-0143-6
[10] ModelBased Systems Engineering (MBSE) Initiative International Council on Systems Engineering (INCOSE), "Survey of Model-Based Systems Engineering (MBSE) Methodologies," INCOSE-TD-2007-003-02, 5/23/2008.
[11] Yue Cao, Yusheng Liu, Christiaan J.J. Paredis, "System-level model integration of design and simulation for mechatronic systems based on SysML", Mechatronics 21 (2011) 1063–1075.
[12] RS Peak, RM Burkhart, SA Friedenthal, MW Wilson, M Bajaj, I Kim, "Simulation-Based Design Using SysML—Part 1: A Parametrics Primer." INCOSE Intl. Symposium, San Diego, 2007.
[13] A. Ramos, J. Ferreira, and J. Barcelo, "Model-Based Systems Engineering: An Emerging Approach for Modern Systems", IEEE Trans on Systems, Man and Cybernetics – Part C: Applications and reviws, Vol. 42, No. 1, Jan 2012, pp. 101-111.
[14] Luca Bassi, Cristian Secchi, Marcello Bonfe, and Cesare Fantuzzi, "A SysML-Based Methodology for Manufacturing Machinery Modeling and Design" IEEE/ASME Trans. On Mechatronics, Vol. 16, No. 6, December 2011, pp. 1049-1062.